\documentclass[twocolumn,prb]{revtex4}
\usepackage[english]{babel}
\usepackage[utf8]{inputenc}
\usepackage{amssymb}
\usepackage{epsfig}

\begin{document}

\newcommand{\de}{\mathrm{d}}
\newcommand{\ak}{a^\dagger}
\newcommand{\h}{\hslash}
\newcommand{\ee}{\varepsilon}
\newcommand{\beq}{\begin{eqnarray}}
\newcommand{\eeq}{\end{eqnarray}}
\newcommand{\is}{& = &}
\newcommand{\nn}{\nonumber}
\newcommand{\la}{\langle}
\newcommand{\ra}{\rangle}
\newcommand{\ie}{{\it i.e.} }
\newcommand{\eg}{{\it e.g.} }

\title%[Spontaneous emission]
      {A simple quantum model of spontaneous emission}
\author{Krzysztof P W\'ojcik} 
\email{kwojcik@hoth.amu.edu.pl}
\affiliation{Faculty of Physics, Adam Mickiewicz University, 
	     61-614 Pozna\'{n}, Poland}
\date{March 31, 2012}

\begin{abstract}

We present a very simple model of a spontaneous emission from a two-level atom,
interacting with a field of a finite number of states. Such a process is often 
said to occur because of the large number of equally-probable states of environment.
We show that in our model increasing the number of field states may and may not 
cause a practically permanent emission, depending on the details of the model. 
We also describe how irreversibility emerges with growing number of states. 
Mathematical tools are reduced to a necessary minimum and hopefully can be well
understood by undergraduate students. 

\end{abstract}

%\submitto{Physics Education}
\maketitle

%%%%%%%%%%%%%%%%%%%%%%
\section{Introduction}
%%%%%%%%%%%%%%%%%%%%%%

In the introductory course on classical thermodynamics, statistical 
thermodynamics is sometimes based on the concept of equally-probable 
microstates\cite{ST}. However, the microscopic mechanism which guarantees 
this equal probability is not discussed. 

Following this path, spontaneous emission may be said to occur because there
exists large (or even infinite) number of excited states of environment and 
only one excited state of the atom. It can be argued that all those states 
are equally probable and thus the equilibrium state is the atom in ground 
state and somehow excited environment. After presenting the model in Section~\ref{sec:building}, in Section~\ref{sec:destroying} we present an 
example of a system, where the permanent emission is absent no matter how 
many states of the environment there are. 

On the other hand, in the typical academic course on quantum mechanics
\cite{QM}, stationary problems constitute a major part.
The only three places where dynamics appear are the definition of time 
evolution operator, description of two-state system and the time-dependent 
perturbation theory. The last is usually presented in a general way for 
arbitrary number of states. The example in which one could trace how the
system properties change with varying number of states is missing. We give
such an example in Section~\ref{sec:emission}.

Thanks to the simplicity of the model, the approach is free of semi-classical 
arguments and does not involve approximate methods like the perturbation theory. 
Thus, the model illustrates how fundamental principles of quantum dynamics works 
in a system more complex then a two-level one and can become a part of an introductory
course on quantum mechanics, making it a little more oriented on dynamics than 
usual. Intuition build on this model may be of a great value when learning more 
advanced topics, like open quantum systems or quantum field theory, where the 
huge or infinite number of states is an essential difficulty.

%%%%%%%%%%%%%%%%%%%%%%%%%%%%
\section{Building the model}
\label{sec:building}
%%%%%%%%%%%%%%%%%%%%%%%%%%%%

We assume that at the initial moment, the system contains single atom in 
excited state and the environment in the ground state. By spontaneous emission 
we understand the following situation: 
{\it We prepare the atom in an excited state. We wait some time $t_w$ and 
 measure if the atom is still excited. We repeat this procedure many times 
 for similar, but not identical $t_w$. We find that if $t_w$ is long enough,
 the atom is usually de-excited.} 
Note, that this definition suits to the experimental reality. In particular,
$t_w$ is not the same in any two real measurements is obvious.

We number all (orthogonal) basis states of the system by $|0\ra, |1\ra, 
\ldots, |N\ra$. We assume that $N$ is finite. Thus, the space of states is an
$(N+1)$-dimensional vector space and the Hamiltonian is an $(N+1)\times (N+1)$
Hermitian matrix (Hermiticity is required for the total probability to be 
conserved during time evolution). 

We choose $|0\ra$ to denote the initial state (\ie the state in which the 
atom is excited and the environment is in its ground state). The states 
$|k\ra$  with $k>0$ correspond to different types of excitation of environment. 
These types of excitations we call ''modes'', in reference to quantum optics. 
Notice, that all basis states are the states of the whole system, \ie the 
atom and the environment. They differ in the part of the system, which is excited.

Now we would like to write down the Hamiltonian of the system, which allows 
the system to flow from the state $|0\ra$ to some state $|k\ra$. In quantum 
mechanics time evolution of the system is described by the time-evolution 
operator $U_t$,
\beq
|\psi(t)\ra \is U_t|\psi(0)\ra = e^{iHt/\h}|\psi(0)\ra.
\label{U}
\eeq 
With the spectral theorem, we immediately see that in the eigenbasis of $H$
the operator $U_t$ is a diagonal matrix with matrix elements
\beq
\la e_n |U_t| e_{n'} \ra \is \delta_{nn' }e^{i E_n t/\h},
\label{exp}
\eeq
where $n,n' \in\{0,1,\ldots, N\}$, $|e_n\ra$ are eigenstates of $H$,
$E_n$ are corresponding eigenvalues and $\delta$ denotes the Kronecker 
delta. In arbitrary basis for small $\Delta t$ we have
\beq
U_{\Delta t} &\approx& I + iH\Delta t/\h
\eeq
and we see that if the Hamiltonian matrix element $\la k |H| 0 \ra$ is 
non-zero, $|\psi(t + \Delta t)\ra$ has different component $|k\ra$ then 
$|\psi(t)\ra$. Otherwise, $\la k |\psi(t + \Delta t)\ra = \la k |\psi(t)\ra$. 
This is valid for all $t$, so $\la k |H| 0 \ra$ must be non-zero if evolution
from $|0\ra$ directly to $|k\ra$ is possible. We denote $\la k |H| 0 \ra = \alpha_k$. 
Due to Hermiticity, also $\la 0 |H| k\ra = \alpha_k^*$ must be non-zero, so 
the reverse process is also possible. This is true for all $k>0$.

We assume that different modes are independent, \ie the Hamiltonian does 
not mix them, $\la k |H| k' \ra = 0$ for $k\neq k'$. This assumption is 
not necessary, but simplifies the model. We denote $\la k |H| k \ra$ 
($k\in \{0,1,\ldots, N\}$) by $\ee_k$.

The probability that the atom is excited after time $t$ is 
$P(t)=|\la 0 | \psi(t) \ra|^2$. Using Eq.~(\ref{U}) and (\ref{exp}), we can 
write it as
\beq
P(t) \is \left| \sum_{n=0}^N |\la 0 | e_n \ra|^2 e^{i E_n t/\h} \right|^2.
\label{P}
\eeq

%%%%%%%%%%%%%%%%%%%%%%%%%%%%%%%%
\section{Destroying stereotypes}
\label{sec:destroying}
%%%%%%%%%%%%%%%%%%%%%%%%%%%%%%%%

Now we would like to describe with our model two systems, where the spontaneous 
emission is not present. They are counterexamples for some widespread stereotypes.

The first example is the case $N=1$ (there is only one excited state of 
environment), $\ee_1=\ee_0$ (mean energies of the state with excited atom and
of the state with excited environment are the same) and arbitrary 
$\alpha_1=\alpha$. Then $H$ is $2\times 2$ matrix,
\beq
H_{1} \is \ee_0 ( |0\ra \la 0| + |1\ra \la 1|) 
	  + \alpha |0\ra\la 1| + \alpha^* |1\ra\la 0|\\
\is \left[ \begin{array}{cc}
	\ee_0  & \alpha\\
	\alpha^* & \ee_0
	\end{array} \right] .\nn
\eeq
It is easy to compute eigenvalues and eigenstates from characteristic 
equation. Then from Eq.~(\ref{P}) follows
\beq
P_{1}(t) \is \cos^2 \left( {|\alpha| \over \h}t \right).
\eeq
This mean that no spontaneous emission is present. The atom
oscillates between the ground and the excited states. We can say it is at 
resonance with the field. Students can individually check that similar 
result holds for arbitrary $\ee_1$ (\ie out of resonance), but the amplitude
of oscillation is smaller and the atom is never certainly de-excited. 

This example is a special case of the familiar 2-level system. It destroys 
the stereotype, that "going to lower energy" is prescribed in any quantum 
theory.

As the second example we consider the case of $N$ identical states, 
\ie $\ee_k=\ee_0$ and $\alpha_k = \alpha$, for some given $N$. The 
Hamiltonian now has the form
\beq
H_{2} \is \ee_0 \sum_{k=0}^N |k\ra \la k|+ 
	  \sum_{k=1}^N (\alpha |0\ra\la k| + \alpha^*|k\ra\la 0|).
\eeq
This example can be solved analytically by a simple change of the basis. 
We denote 
\beq
|\tilde 1\ra \is {1\over \sqrt{N} } \sum_{k=1}^N |k\ra.
\label{one}
\eeq
The new basis consists of $|\tilde 0\ra=|0\ra, |\tilde 1\ra,$ and any orthogonal to
them (and each other) and normalized states $|\tilde 2 \ra, \ldots, |\tilde N\ra$. 
The part of Hamiltonian proportional to the identity matrix $\sum_{k=0}^N |k\ra \la k|$
looks the same in any basis. Using Eq.~(\ref{one}) and the facts that 
$\la 0|\tilde k \ra = 0$ is the condition of orthogonality to $|\tilde 0\ra$ and 
$\sum_{k=1}^N \la k | \tilde k \ra = 0$ is the condition of orthogonality to 
$|\tilde 1\ra$, the following part can be expressed as
\beq
\sum_{k=1}^N \alpha | 0 \ra \la k | +h.c. 
	\is \sum_{k=1}^N \sum_{\tilde k=\tilde 0}^{\tilde N} \alpha 
	    | 0 \ra \la k | \tilde k \ra\la \tilde k| + h.c.\\
	\is   \sqrt{N} \alpha |0\ra \la \tilde 1| +h.c. \nn
\label{H}
\eeq
($h.c.$ denotes "Hermitian conjugate"). Thus all states orthogonal to 
$|\tilde 0\ra$ and $|\tilde 1\ra$ are eigenstates with eigenvalue $\ee_0$. 
This means that we only have to find two last eigenstates, \ie 
diagonalize remaining $2\times 2$ submatrix. Further calculation and the 
resulting evolution is similar to the case $N=1$. The probability that the 
atom is excited is
\beq
P_{2}(t) \is \cos^2 \left( {\sqrt{N} |\alpha| \over \h}t \right).
\label{P2}
\eeq
No spontaneous emission is present, no matter how large $N$ is. This example
destroys the stereotype that microstates are always equally probable and the 
huge number of possible states affirm that spontaneous emission is present 
in the system.

%%%%%%%%%%%%%%%%%%%%%%%%%%%%%%%%%%%%%%%%%%%%%
\section{Conditions for spontaneous emission}
\label{sec:emission}
%%%%%%%%%%%%%%%%%%%%%%%%%%%%%%%%%%%%%%%%%%%%%

To understand when spontaneous emission occurs we analyse Eq.~(\ref{P}).
We notice that the expression for $P(t)$ resembles a bit a square of module
of some Fourier series. Consider the special case when $N=2M$. To parametrize
eigenvalues and eigenstates we use $m \in \{-M, \ldots, M\}$ instead of 
$n\in \{0,\ldots 2M\}$. We assume that the eigenvalues are equally-spaced
in the section $\ee_0 \pm D$, 
\beq
E_m \is \ee_0 + {m \over M} D,
\label{Em}
\eeq
and that
\beq
\la 0 | e_m\ra \is \la 0 |e_{-m}\ra.
\eeq
We would like to stress, that now we make assumptions about the eigenvalues
of the whole system and we have not yet discussed the Hamiltonian matrix 
elements $\ee_k$ and $\alpha_k$ in our initial basis.

From Eq.~(\ref{P}) we have
\beq
\label{Fourier}
\sqrt{P_3(t)} \is \left| e^{i \ee_0 t/\h} \sum_{m=-M}^M |\la 0 | \tilde e_m \ra|^2
		  e^{i {m \over M} D t/\h} \right|\\
	      \is \left| |\la 0 | \tilde e_0 \ra|^2 + \sum_{m=1}^M 
		  2|\la 0 | \tilde e_m \ra|^2 \cos 
		  \left(i {m \over M} D t/\h \right) \right|.\nn
\eeq
Eq.~(\ref{Fourier}) states that $\sqrt{P_3(t)}$ can be expand in a (finite) 
Fourier series. Thus it is a periodic function with period 
\beq
T = 2\pi\h {M \over D} = h {M \over D}.
\eeq
However, this does not mean that spontaneous emission is not present in the
system. First, in reality the dimension of space of states is surely huge, 
possibly infinite. For large $M$, $T$ is also large. There exists such a big 
$M$ that $T$ is too long to be ever measured. What is more, even when $T$ 
is quite small, $\sqrt{P(t)}$ may be measurably non-zero only for a very 
short time $\tau$ in every period. If unavoidable differences in $t_w$ in 
different measurements are significantly larger then $\tau$, spontaneous 
emission is observed.

Note that the coefficients of Fourier expansion are dependent only on 
scalar products of initial state $|0\ra$ with eigenstates $|\tilde e_m\ra$. The 
only properties that they must fulfil for all choices of $\ee_k$ and 
$\alpha_k$ are that they are positive real numbers summing to the unity. To 
prove this we show the algorithm which finds $\ee_k$ and $\alpha_k$:
  \begin{enumerate}
  \item Choose $|\la 0 | \tilde e_m \ra|^2 > 0$. $E_m$ fulfils Eq.~(\ref{Em}). 
	In its eigenbasis, $H$ is a diagonal matrix with elements $E_m$. 
  \item $|0\ra$ may be chosen to have real, positive components in 
	the eigenbasis of $H$. Their moduli are determined by the choice of 
	$|\la 0 | \tilde e_m \ra|^2$.
  \item Orthonormalize $\{|e_{-M}\ra, \ldots, |e_{-1}\ra, |0\ra, |e_{1}\ra,
	\ldots, |e_{M}\ra \}$ using Gram-Schmidt procedure, starting from 
	$|0\ra$. Find $H$ in this basis.
  \item Consider the subspace orthogonal to $|0\ra$. Denote $H$ reduced to 
	this subspace by $H_{red}$. Find the eigenbasis of $H_{red}$.
  \item Change the basis such, that $|0\ra$ stays invariant and the 
	following basis states are eigenstates of $H_{red}$ extended to the 
	original space. Calculate $H$ in this basis.
  \item The result is the Hamiltonian matrix which has the desired form.
  \end{enumerate}
Realizing this algorithm numerically for some concrete $|\la 0 | \tilde e_m \ra|$ 
(\eg in the example considered below) may be an instructive exercise for students.

The corollary is that there exist very wide range of possible $P(t)$. For 
example, we can choose
\beq
|\la 0 | e_m \ra|^2 \is {1 \over 2M+1}
\label{ck}
\eeq
for any $m$. This is not a blind guess. We know, that the Fourier transform 
of the Dirac delta function is constant, so we expect that with Eq.~(\ref{ck}) 
$\sqrt{P_3(t)}$ will be considerably different from zero only for $t$ being 
close to integer multiple of $T$. The results are plotted in Figure 1. It 
can be seen, how period becomes larger when $M$ increases. Simultaneously,
 the time after which the emission happens is approximately independent of 
$M$. Thus for big enough (but still finite) $M$, the probability that after 
long and not precisely determined time $t_w$ the atom is excited is not 
measurable.

\begin{figure}[ht]
\centering
\includegraphics[width=0.5\textwidth]{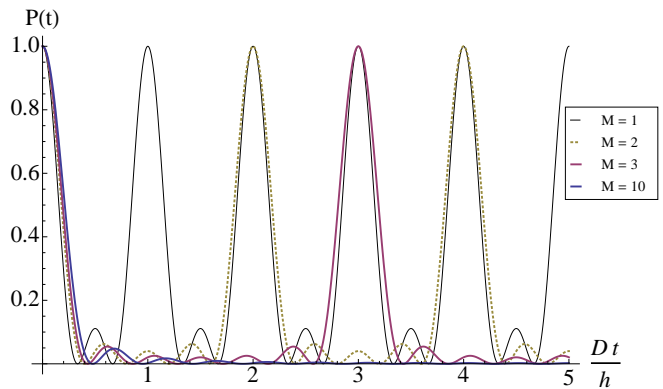}
\caption{Probability $P(t)$ that the atom is in excited state after time $t$
	 when the relation of the initial state to eigenstates is given by 
	 Eq.~(\ref{ck}) and the energy eigenvalues are given by Eq.~(\ref{Em}). 
	 Total number of states equals $2M+1$.}
\end{figure}

%%%%%%%%%%%%%%%%%%%%%%%%%%%%%%%%%%
\section{Conclusions of the model}
\label{sec:conclusions}
%%%%%%%%%%%%%%%%%%%%%%%%%%%%%%%%%%

We propose a simple quantum model for the system consisted of the two-level
atom and $N$-modal field. We have shown, that large number of modes is a
necessary but (contrary to widespread opinion) not sufficient condition for
occurrence of spontaneous emission. Whether the emission is present in the 
system or not, depends on the interaction with the field and the modes mean
energies, which determine the coefficients $|\la 0 | \tilde e_m \ra|^2$ in
Eq.~(\ref{P}).

We have shown the algorithm of building a Hamiltonian which cause $P(t)$ to 
be desired function if only square root of this function has a finite 
expansion in the cosine Fourier series (with any period). This is an extremely 
broad set of functions. This shows that even a simple two-level atom can undergo 
various types of an evolution when is placed in a complex environment. 

\section{Acknowledgements}

I would like to thank all the participants of the 6$^{\rm th}$ Scientific 
Camp of Physics Students Club from University of Warsaw, which took place
in Bukowina Tatrza{\'n}ska in Poland \hbox{17-24.07.2011}, for inviting me
to the Camp and for the discussions on the early version of the model and 
critique of it. I am also grateful to R. Chhajlany for all kinds of help.

\end{document}